\documentclass[10pt,twoside]{article}

\usepackage{graphicx}
\usepackage{amssymb}
\usepackage{latexsym}

\makeatletter

\@addtoreset{equation}{section} \makeatother
\newtheorem{theorem}{Theorem}[section]
\newtheorem{remark}[theorem]{Remark}
\newtheorem{lemma}[theorem]{Lemma}
\newtheorem{proposition}[theorem]{Proposition}

\newtheorem{definition}[theorem]{Definition}

\newcommand{\be}{\begin{equation}}
\newcommand{\ee}{\end{equation}}

\font\ddpp=msbm10 scaled \magstep 1

\def\H{\hbox{\ddpp H}}

\newcommand{\R}{{\mathbb R}}
\newcommand{\e}{{\mathbb S}}

\title{{\bf The causal boundary of product spacetimes}}

\author{V. ALA\~NA, J.L. FLORES\thanks{The second
author has been supported by MEC Grants RyC-2004-382 and
MTM-2004-06262.}
\\
{\small Departamento de \'Algebra, Geometr\'{\i}a y Topolog\'{\i}a}\\
{\small Facultad de Ciencias, Universidad de M\'alaga}\\
{\small Campus Teatinos, 29071 M\'alaga, Spain}\\
{\small E-mail: floresj@agt.cie.uma.es}\\ {\small Phone/Fax:
(+34)952132387/2008}
\\}

\begin{document}
\parindent=5mm
\date{}
\maketitle

\begin{quote}

\noindent {\small \bf Abstract.}

{\small The new formulation of the causal completion of spacetimes
suggested in \cite{MR}, and modified later in \cite{F}, is tested by
computing the causal boundary for product spacetimes of a Lorentz
interval and a Riemannian manifold. This is particularized for two
important families of spacetimes, conformal to the previous ones:
(standard) static spacetimes and Generalized Robertson-Walker
spacetimes. As consequence, it is shown that this new approach
essentially reproduces the structure of the conformal boundary for
multiple classical spacetimes: Reissner-Nordstrom (including
Schwarzschild), Anti-de Sitter, Taub and standard cosmological
models as de Sitter and Einstein Universe.}
\end{quote}
\begin{quote}
\end{quote}

\newpage
%--------------------------------------------------------------

\section{Introduction}

In the last decades, relativists have shown great interest in
certain remarkable properties related to the asymptotic behavior of
spacetimes, as singularities. In order to get a better understanding
of these phenomenons, sometimes it is very useful to attach a sort
of ideal boundary to the spacetime. However, the construction of the
`optimal' boundary has shown to be a very elusive problem up to
date.

A well-known boundary in Relativity is the {\em conformal boundary}
\cite{P}, which consists of conformally embedding the original
spacetime into a larger one, and then, taking the boundary of the
image. However, this method is neither systematic nor intrinsic, and
sometimes it results very restrictive. In order to overcome these
handicaps, Geroch, Kronheimer and Penrose introduced a new
construction called {\em causal boundary} \cite{GKP}. In this new
approach they attach a future (past) ideal point for every
inextensible, physically admissible future (past) trajectory, in
such a way that the ideal point only depends on the past (future) of
the trajectory. In fact, this method is systematic, intrinsic and
very general. However, it suffers from an important technical
difficulty: in general, some future and past ideal points must be
identified in order to avoid pathologies derived from having `too
big' boundaries.

Many authors have tried different methods to establish these
identifications \cite{GKP,BS,R,S}; indeed, this question is related
to the introduction of a satisfactory topology for the completion.
However, they have not obtained totally satisfactory results up to
date (see \cite{KLL,KL1,KL2,MR}). See also \cite{GS2}, \cite{H5} for
interesting reviews on the subject.

Based on a new formulation by Marolf and Ross \cite{MR}, which
replace the identifications by pairs representing the ideal points,
recently the second author has developed a new approach to the
causal boundary with promising results \cite{F}. This approach lies
on a `minimality principle' which allows to establish the desired
pairs, in addition to a reasonable topology for the completion. As
consequence, it is obtained a construction with many satisfactory
mathematical properties. However, this construction has not been
checked in many spacetimes of physical interest.

In this paper we are going to test this new formulation by computing
the causal boundary for some physically relevant spacetimes. After a
preliminary section devoted to recall some basic notions on causal
structure and causal completions, in Section \ref{3} we construct
the causal boundary for product spacetimes of a Lorentz interval and
a Riemannian manifold. In Section \ref{4} we use the conformal
invariance of the causal boundary to directly deduce the boundary
for two important families in the conformal class: (standard) static
spacetimes and Generalized Robertson-Walker spacetimes. As
consequence, we describe the boundary for multiple classical
spacetimes in these families: Reissner-Nordstrom (including
Schwarzschild), Anti-de Sitter, Taub and standard cosmological
models as de Sitter and Einstein Universe. Finally, in Section
\ref{5} we summarize the main conclusions, putting special emphasis
in the fact that this approach essentially reproduces the structure
of the conformal boundary for these examples.

\section{Preliminaries}\label{preliminaries}

Let $(V,g)$ be a spacetime, i.e. a connected smooth manifold $V$
endowed with metric tensor $g$ of index $1$. A tangent vector $v\in
T_{p}V$, $p\in V$ is named {\em timelike} (resp. {\em lightlike};
{\em spacelike}; {\em causal}) if $g(v,v)<0$ (resp. $g(v,v)=0$ and
$v\neq 0$; $g(v,v)>0$ or $v=0$; $v$ is either timelike or
lightlike). Accordingly, a smooth curve $\gamma:I\rightarrow V$ ($I$
real interval) is called {\em timelike} (resp. {\em lightlike}; {\em
spacelike}; {\em causal}) if $\gamma'(s)$ is timelike (resp.
lightlike; spacelike; causal) for all $s$. Spacetimes are assumed to
be {\em time-orientable}, i.e. they must admit a {\em
time-orientation}, which is a continuous, globally defined, timelike
vector field $X$. Fixed a time-orientation $X$, a causal curve
$\gamma(s)$ is said {\em future-directed} (resp. {\em
past-directed}) if $g(\gamma'(s),X(\gamma(s)))<0$ (resp.
$g(\gamma'(s),X(\gamma(s)))>0$) for all $s$. Future-directed causal
curves represent all the physically admissible trajectories for
material particles and light rays in the universe.

Two events $p,q\in V$ are {\em chronologically related} $p\ll q$
(resp. {\em causally related} $p\prec q$) if there exists some
future-directed timelike (resp. causal) curve from $p$ to $q$. If
$p\prec q$ but $p\not\ll q$, they are said {\em horismotically
related} $p\rightarrow q$. The {\em chronological past} of $p$,
$I^{-}(p)$, (resp. {\em causal past} of $p$, $J^{-}(p)$) is defined
as:
\[
I^{-}(p)=\{q\in V: q\ll p\}\qquad(\hbox{resp.}\;\; J^{-}(p)=\{q\in
V: q\prec p\}).
\]
Of course, the {\em chronological future} of $p$, $I^{+}(p)$ (resp.
{\em causal future} of $p$, $J^{+}(p)$) is defined by replacing
$q\ll p$ (resp. $q\prec p$) by $p\ll q$ (resp. $p\prec q$) in
previous definition.

The main purpose of the causal completion of spacetimes is to avoid
the existence of inextensible timelike curves. This is overcome by
adding {\em ideal points} to the spacetime in such a way that any
timelike curve presents some endpoint in the new space. In order to
rigorously describe this completion, applicable to {\em strongly
causal} spacetimes (i.e. spacetimes without closed or `nearly
closed' timelike curves), previously we need to introduce some
terminology:

\noindent A subset $P\subset V$ is called {\em past set} if it
coincides with its past, which is always open; i.e.
$P=I^{-}[P]:=\{p\in V: p\ll q\;\hbox{for some}\; q\in P\}$. Given a
subset $S\subset V$, the {\em common past} of $S$ is defined by
$\downarrow S:=I^{-}[\{p\in V:\;\; p\ll q\;\;\forall q\in S\}]$. A
past set that cannot be written as the union of two proper subsets,
both of which are also past sets, is called {\em indecomposable
past} set, IP. An IP which does not coincide with the past of any
point in $V$ is called {\em terminal indecomposable past set}, TIP.
Otherwise, it is called {\em proper indecomposable past set}, PIP.
By interchanging the roles of past and future, we obtain the
corresponding notions for {\em future set}, {\em common future}, IF,
TIF and PIF.

In order to construct the future causal completion, first identify
every event $p\in V$ with its PIP, $I^{-}(p)$. Then, the {\em
future causal boundary} $\hat{\partial}(V)$ of $V$ is defined as
the set of all TIPs in $V$. Therefore, {\em the future causal
completion} $\hat{V}$ becomes the set of all IPs:
\[
V\equiv \hbox{PIPs},\qquad \hat{\partial}(V)\equiv
\hbox{TIPs},\qquad\hat{V}\equiv \hbox{IPs}.
\]
Analogously, every event $p\in V$ can be identified with its PIF,
$I^{+}(p)$. Then, the {\em past causal boundary}
$\check{\partial}(V)$ of $V$ is defined as the set of all TIFs in
$V$, and thus, {\em the past causal completion} $\check{V}$ is the
set of all IFs:
\[
V\equiv \hbox{PIFs},\qquad \check{\partial}(V)\equiv
\hbox{TIFs},\qquad\check{V}\equiv \hbox{IFs}.
\]

For the (total) causal completion, one immediately thinks of the
space $\hat{V}\cup\check{V}$. However, it becomes evident that by
only imposing the obvious identifications $I^{-}(p)\sim I^{+}(p)$ on
$\hat{V}\cup\check{V}$ for all $p\in V$, the resulting space
$V^{\natural}$ does not provide a satisfactory description for the
boundary of $V$: in fact, this procedure often attaches two ideal
points where we would expect only one.

The first attempt to establish identifications in
$\hat{\partial}(V)\cup\check{\partial}(V)$ was proposed in
\cite{GKP}. The authors introduced a generalized Alexandrov topology
on $V^{\natural}$: the topology  generated by the sub-basis
\[
\begin{array}{l}
A^{int}=\{P\in\hat{V}: P\cap A\neq\emptyset\} \\
A^{ext}=\{P\in\hat{V}: P=I^{-}[W]\;\;\hbox{implies}\;\;
I^{+}[W]\nsubseteq A\}
\end{array}\qquad\hbox{for all}\;\; A\in \check{V},
\]
\[
\begin{array}{l}
B^{int}=\{F\in\check{V}: F\cap B\neq\emptyset\} \\
B^{ext}=\{F\in\check{V}: F=I^{+}[W]\;\;\hbox{implies}\;\;
I^{-}[W]\nsubseteq B\}
\end{array}\qquad\hbox{for all}\;\; B\in \hat{V}.
\]
Then, they suggested the minimum set of identifications necessary to
obtain a Hausdorff space. However, this method fails to produce the
`expected' structure and topology for the completion in some
examples \cite{KLL}, \cite{KL1}, \cite{H2}, \cite[Sect. 5]{MR}. As
commented before, other more accurate attempts have been suggested
since then, but without totally satisfactory results.

An alternative procedure to making identifications consists of
forming pairs composed by past and future indecomposable sets of
$V$. This approach, firstly introduced in \cite{MR} and developed
later in \cite{F}, has exhibited satisfactory results for those
spacetimes analyzed up to date (see \cite{MR,F,FS}). In this paper
we are going to test this approach for product spacetimes.

Even if the criteria proposed in \cite{MR} and \cite{F} for pairing
terminal sets are different in general, they coincide in many cases.
In particular, they coincide for those spacetimes such that every
terminal set is not S-related (Szabados related) with more than one
terminal set: we say that $P,F$ are {\em S-related}, $P\sim_{S}F$,
if $P$ is maximal IP into $\downarrow F$ and $F$ is maximal IF into
$\uparrow P$. For these spacetimes, the construction in \cite[Th.
7.4]{F} reduces to the following definition coming from \cite[Def.
4]{MR}:
\begin{definition}\label{d} The {\em (total) causal boundary} $\overline{V}$ is formed by all
the pairs $(P,F)$ formed by a TIP $P$ and a TIF $F$ such that:
either $P\sim_{S}F$; or $P=\emptyset$ and there is no $P'\neq
\emptyset$ such that $P'\sim_{S}F$; or $F=\emptyset$ and there is no
$F'\neq\emptyset$ such that $P\sim_{S}F'$.
\end{definition}
This will be the definition adopted in this paper, since product
spacetimes always satisfy the property above (Remark \ref{r}).
%\footnote{In fact, Definition \ref{d} is a special case of the more
%general construction in [F], named {\em chronological completion}.}.

With this definition, the causal structure of the spacetime can be
easily extended to the completion. Concretely, we say that $(P,F),
(P',F')\in \overline{V}$ are {\em chronologically related},
$(P,F)\ll (P',F')$, if $F\cap P'\neq\emptyset$. Here, $(P,F)$,
$(P',F')$ will be said {\em causally related}, $(P,F)\prec (P',F')$,
if $F'\subset F$ and $P\subset P'$. Finally, $(P,F)$, $(P',F')$ are
{\em horismotically related}, $(P,F)\rightarrow (P',F')$, if they
are causally, but not chronologically, related.

The topology of the spacetime can be also extended to the
completion. Here we adopt the so called {\em chronological
topology}, firstly introduced in \cite{F}. This topology is defined
in terms of the following {\em limit operator} $L$: given a sequence
$\sigma=\{(P_{n},F_{n})\}\subset\overline{V}$, we say that $(P,F)\in
L(\sigma)$ if\footnote{By LI and LS we must understand the usual
inferior and superior limits of sets: i.e. LI$(A_{n})\equiv
\liminf(A_{n}):=\cup_{n=1}^{\infty}\cap_{k=n}^{\infty}A_{k}$ and
LS$(A_{n})\equiv
\limsup(A_{n}):=\cap_{n=1}^{\infty}\cup_{k=n}^{\infty}A_{k}$.}
\begin{equation}\label{ee}
\begin{array}{c}
P\in\hat{L}(P_{n}):=\{P'\in\hat{V}: P'\subset {\rm
LI}(P_{n})\;\;\hbox{and}\;\; P'\;\;\hbox{is maximal IP into}\;\; {\rm LS}(P_{n})\} \\
F\in\check{L}(F_{n}):=\{F'\in\check{V}: F'\subset {\rm
LI}(F_{n})\;\;\hbox{and}\;\; F'\;\;\hbox{is maximal IF into}\;\;
{\rm LS}(F_{n})\}.
\end{array}
\end{equation}
Then, the {\em closed sets} for the chronological topology are those
subsets $C\subset \overline{V}$ such that $L(\sigma)\subset C$ for
any sequence $\sigma\subset C$.

Finally, we remark that the causal boundary is conformally
invariant, i.e. it remains unaltered under conformal transformations
of the spacetime.

\section{Causal boundary for product spacetimes}\label{3}

We consider {\em product spacetimes} $(V,g)$ of a Lorentz interval
$(I,-dt^{2})$, $I=(a,b)\subseteq\R$ and a Riemannian manifold
$(M,h)$:
\begin{equation}\label{uhh}
V=I\times M,\qquad g=-dt^{2}+h.
\end{equation}
Here, the time orientation is determined by $\partial_{t}$. Of
course, these spacetimes are always strongly causal, and thus, the
causal completion applies. In order to construct the causal boundary
of these spacetimes, we have followed several steps. By
completeness, we have included here some arguments and results
coming from \cite{H3}.

\vspace{3mm}

\noindent \textbf{\em $\S$1. Computation of PIPs and PIFs:}

\vspace{2mm}

We begin by computing the PIPs and PIFs. If $(t_{0},x_{0})\ll
(t_{1},x_{1})$ then there exists a timelike curve
$\gamma(s)=(t(s),c(s))$ such that $\gamma(0)=(t_{0},x_{0})$,
$\gamma(1)=(t_{1},x_{1})$. Since $\gamma$ is timelike, necessarily
$\dot{t}(s)>\sqrt{h(\dot{c}(s),\dot{c}(s))}$ for all $s$. In
particular,
\[
t_{1}-t_{0}=\int_{0}^{1}\dot{t}(s)ds>\int_{0}^{1}\sqrt{h(\dot{c}(s),\dot{c}(s))}=\hbox{long}(c)\geq
d(x_{0},x_{1}),
\]
being $d$ the distance in $M$ associated to $h$. Reciprocally, if
$t_{1}-t_{0}>d(x_{0},x_{1})$ then there exists a curve $c(s)$ in $M$
with $c(0)=x_{0}$, $c(1)=x_{1}$ such that $|\dot{c}(s)|<t_{1}-t_{0}$
for all $s$. Therefore, the curve $\gamma(s)=(t(s),c(s))$, with
$t(s)=(t_{1}-t_{0})s+t_{0}$, is timelike and satisfies
$\gamma(0)=(t_{0},x_{0})$, $\gamma(1)=(t_{1},x_{1})$, proving that
$(t_{0},x_{0})\ll (t_{1},x_{1})$. Summarizing:
\begin{equation}\label{marras}
(t_{0},x_{0})\ll (t_{1},x_{1})\Longleftrightarrow
t_{0}<t_{1}-d(x_{0},x_{1}).
\end{equation}
This property directly provides the following result:
\begin{proposition} The PIPs and PIFs of product spacetime (\ref{uhh}) are:
\begin{equation}\label{vv}
\begin{array}{c}
I^{-}(p)=\{(t',x')\in I\times M:
t'<d^{t}_{x}(x')\} \\
I^{+}(p)=\{(t',x')\in I\times M: t'>-d^{-t}_{x}(x')\}
\end{array}
\quad\hbox{for all}\;\; p=(t,x)\in I\times M,
\end{equation}
being $d^{t}_{x}(\cdot):=t-d(\cdot,x)$.
\end{proposition}

\vspace{3mm}

\noindent \textbf{\em $\S$2. Computation of TIPs and TIFs:}

\vspace{3mm}

To this aim, we only need to compute the past and future of
inextensible timelike curves (see, for example, \cite[Prop.
6.14]{BEE}). So, let $\gamma(s)=(t(s),c(s))$ be an inextensible
future-directed timelike curve. In particular, $\dot{t}(s)>0$ for
all $s$, and thus, we can reparametrize $\gamma$ by $t$ in order to
obtain $\gamma(t)=(t,c(t))$, where now the spatial projection $c$ is
a curve with domain some interval $[w,\Omega)$, $\Omega\leq b$ and
velocity $|\dot{c}|<1$. The following definition will be useful:
\begin{definition} We define the {\em Busemann function} of such a curve $c$ as the function:
\[
b_{c}:M\rightarrow\R^{*}\equiv\R\cup\{\infty\},\quad
b_{c}(\cdot):=\lim_{t\rightarrow\Omega}d^{t}_{c(t)}(\cdot)=\lim_{t\rightarrow\Omega}(t-d(\cdot,c(t))).
\]
\end{definition}
Recall that the past of $\gamma$ coincides with the union of the
pasts $I^{-}(\gamma(t))$ $\,\forall t\in [w,\Omega)$. Therefore,
$(t',x')\in I^{-}[\gamma]$ if and only if $(t',x')\in
I^{-}(\gamma(t))$ for some $t$ close enough to $\Omega$ (observe
that $I^{-}(\gamma(t_{1}))\subset I^{-}(\gamma(t_{2}))$ if
$t_{1}<t_{2}$). Taking into account (\ref{vv}), this condition
translates into the following inequality:
\[
t'<\lim_{t\rightarrow\Omega}d^{t}_{c(t)}(x')=b_{c}(x').
\]
If $\gamma(t)=(-t,c(t))$ is an inextensible past-directed timelike
curve, analogous arguments can be applied (just interchange the
roles of future and past) in order to obtain: $(t',x')\in
I^{+}[\gamma]$ if and only if
\[
t'>\lim_{t\rightarrow\Omega}(-d^{t}_{c(t)}(x'))=-b_{c}(x').
\]
Summarizing, we can establish the following result:
\begin{proposition}\label{pepe} The TIPs (TIFs) of product
spacetime (\ref{uhh}) are:
\[
\begin{array}{c} I^{-}[\gamma]=\{(t',x')\in I\times M: t'<b_{c}(x')\} \\ (I^{+}[\gamma]=\{(t',x')\in I\times M:
t'>-b_{c}(x')\})
\end{array}
\]
for any inextensible future (past) timelike curve $\gamma$.
\end{proposition}

\vspace{3mm}

\noindent \textbf{\em $\S$3. Partial boundaries:}

\vspace{2mm}

The structure of the partial boundaries can be analyzed in terms of
the extremes of the interval:

\vspace{2mm}

{\bf Case $b=\infty$.} Any inextensible curve $\gamma$ with
component $c$ approaching to some $x_{0}\in M$ satisfy
$\Omega=\infty$. Moreover, in this case $b_{c}\equiv\infty$.
Therefore, from Proposition \ref{pepe} these curves satisfy
\[
I^{-}[\gamma]=V.
\]
This TIP corresponds to the {\em future timelike infinity}, and is
labeled by $i^{+}$. The rest of TIPs are univocally determined by
all the finite Busemann functions $b_{c}<\infty$ in $(M,h)$
associated to inextensible components $c$ of curves $\gamma$.
Therefore, if ${\cal B}(M)$ denotes the set of all these finite
Busemann functions, then we have:
\[
\hat{\partial}(V)={\cal B}(M)\cup\{\infty\}.
\]
The set ${\cal B}(M)$ is invariant under the additive action: if
$b_{c}\in {\cal B}(M)$ then $b_{c}+k\in {\cal B}(M)$ for all $k\in
(a-\Omega,\infty)$ (in fact, $b_{c'}=b_{c}+k$ if $c'(t):=c(t-k)$).
Therefore, if we define the {\em Busemann boundary} as the quotient
\[
\partial_{B}(M):={\cal B}(M)/(a,\infty),
\]
then it is
\[
\hat{\partial}(V)={\cal
B}(M)\cup\{\infty\}\equiv(\partial_{B}(M)\times (a,\infty))\cup
\{i^{+}\}.
\]
It should be remarked that $\partial_{B}(M)$ includes two types of
elements. Those points associated to inextensible curves $c$ with
$\Omega=\infty$, which can be interpreted as `infinity directions'
of the manifold $(M,h)$; and those points associated to inextensible
curves $c$ with $\Omega<\infty$, which define points of the {\em
Cauchy boundary} $\partial_{C}(M)$ of the manifold. In this last
case we have $b_{c}=d_{x_{0}}^{\Omega}$, $x_{0}\in\partial_{C}(M)$.

\vspace{2mm}

{\bf Case $b<\infty$.} Now $\Omega<\infty$, and thus, $b_{c}<\infty$
for any $c$. In particular, $i^{+}$ does not belong to the future
boundary of the spacetime. Indeed, every inextensible curve $\gamma$
with $c$ approaching to some $x_{0}\in M$, and thus, $\Omega=b$, has
Busemann function $b_{c}=d^{b}_{x_{0}}<\infty$. As consequence, the
future boundary contains a copy of $M$. The rest of TIPs are
univocally determined by all the finite Busemann functions
$b_{c}<\infty$ associated to inextensible components $c$ of curves
$\gamma$. We denoted this set by ${\cal B}(M)$. Arguing as before,
${\cal B}(M)$ is invariant by the additive action: if $b_{c}\in
{\cal B}(M)$ then $b_{c}+k\in {\cal B}(M)$ for all $k\in
(a-\Omega,b-\Omega]$. Therefore:
\[
\hat{\partial}(V)={\cal B}(M)\cup M\equiv(\partial_{B}(M)\times
(a,b])\cup M\equiv(\partial_{C}(M)\times (a,b])\cup M.
\]

\vspace{2mm}

If we repeat the arguments above, but now interchanging the roles
of future and past, we obtain the corresponding results for the
past boundaries in terms of the extreme $a$. That is:

\vspace{2mm}

{\bf Case $a=-\infty$.} Now $\check{\partial}(V)={\cal
B}(M)\cup\{-\infty\}\equiv(\partial_{B}(M)\times (-\infty,b))\cup
\{i^{-}\}$, where $i^{-}$ labels the TIF $V$ corresponding to the
{\em past timelike infinity}.

\vspace{2mm} {\bf Case $a>-\infty$.} Now $\check{\partial}(V)={\cal
B}(M)\cup M\equiv(\partial_{C}(M)\times [a,b))\cup M$.

\vspace{3mm}

\noindent \textbf{\em $\S$4. The (total) causal boundary:}

\vspace{2mm}

In order to construct the (total) causal boundary from the partial
boundaries, we need to know which TIPs $P$ and TIFs $F$ are
S-related. The following lemma solves this question:
\begin{lemma}\label{pr} Two terminal sets
$P,F\neq\emptyset$ of $V$ satisfy $P\sim_{S}F$ iff for some
$x_{0}\in\partial_{C}(M)$
\begin{equation}\label{i}
P=\{(t',x'): t'<d^{\Omega}_{x_{0}}(x')\},\quad F=\{(t',x'):
t'>-d^{-\Omega}_{x_{0}}(x')\},\quad a<\Omega<b.
\end{equation}
\end{lemma}
{\it Proof.} Suppose that $P\sim_{S}F$, with $P,F\neq\emptyset$.
Let $\gamma_{+}:[w,\Omega)\rightarrow V$,
$\gamma_{+}(t)=(t,c_{+}(t))$ be an inextensible future-directed
timelike curve such that $P=I^{-}[\gamma_{+}]$. Since $\uparrow
P\supset F\neq\emptyset$ and coordinate $t$ strictly increases
along future-directed timelike curves, necessarily $\Omega<b$.
Moreover, since $\gamma_{+}$ is inextensible, necessarily
$c_{+}(t)\rightarrow x_{0}\in\partial_{C}(M)$. Analogously,
$F=I^{+}[\gamma_{-}]$, where $\gamma_{-}:[w',\Omega')\rightarrow
V$, $\gamma_{-}(t)=(-t+2\Omega',c_{-}(t))$, is an inextensible
past-directed timelike curve such that $c_{-}(t)\rightarrow
x'_{0}\in\partial_{C}(M)$ and $\Omega'>a$. Summarizing:
\[
P=\{(t',x'): t'<d^{\Omega}_{x_{0}}(x')\},\quad F=\{(t',x'):
t'>-d^{-\Omega'}_{x'_{0}}(x')\},
\]
being $a<\Omega, \Omega'<b$ and $x_{0}, x'_{0}\in\partial_{C}(M)$.
Moreover, it is $d(x_{0},x'_{0})\leq \Omega'-\Omega$, since,
otherwise, we can take $\overline{\Omega}'>\Omega'$ and
$\overline{\Omega}<\Omega$ such that
$d(x_{0},x'_{0})>\overline{\Omega}'-\overline{\Omega}$, which
implies $(\overline{\Omega},x_{0})\in P$,
$(\overline{\Omega}',x'_{0})\in F$ but
$(\overline{\Omega},x_{0})\not\ll (\overline{\Omega}',x'_{0})$, in
contradiction with $F\subset\uparrow P$. Finally, notice also that
$\Omega=\Omega'$, and thus, $x_{0}=x'_{0}$. In fact, take any
$x_{0}\in\overline{M}_{C}$ and $\Omega<\overline{\Omega}<\Omega'$
such that $d(x_{0},\overline{x}_{0})=\overline{\Omega}-\Omega$ and
$d(x'_{0},\overline{x}_{0})=\Omega'-\overline{\Omega}$. Then,
$\overline{F}:=\{(t',x'):
t'>-d_{\overline{x}_{0}}^{-\overline{\Omega}}(x')\}$ satisfies
$F\varsubsetneq\overline{F}\subseteq \uparrow P$, which contradicts
the maximality of $F$ into $\uparrow P$.

Assume now that (\ref{i}) holds for some $x_{0}\in\partial_{C}(M)$.
If $(t,x)\in F$ then $t-\Omega>d(x,x_{0})$, which implies $(t,x)\in
\uparrow P$. Therefore, $F\subset \uparrow P$. Moreover, $F$ is
maximal into $\uparrow P$, since, otherwise, there would exist
$(t,x)\neq (\Omega,x_{0})$ satisfying $t-\Omega\geq d(x,x_{0})$ and
$\Omega-t\geq d(x,x_{0})$, a contradiction. Analogously, we can
prove that $P$ is maximal into $\downarrow F$. Whence, $P\sim_{S}F$.
$\Box$

\begin{remark}\label{r} {\em In particular, this shows that every terminal set is not S-related with more
than one terminal set.}
\end{remark}

\vspace{1mm}

\noindent From Definition \ref{d}, Lemma \ref{pr} and Remark
\ref{r}, we deduce the following result:
\begin{proposition} The causal boundary $\partial(V)$ of
$V$ can be written as the union of the corresponding partial
boundaries $\hat{\partial}(V)$, $\check{\partial}(V)$, with each
pair of lines in $\hat{\partial}(V)$, $\check{\partial}(V)$ based on
the same point of $\partial_{C}(M)$ identified.
\end{proposition}

\vspace{3mm}

\noindent \textbf{\em $\S$5. Causal structure and topology for the
boundary:}

\vspace{2mm}

The causal relations between ideal points are given by this
proposition:
\begin{proposition}\label{ant} Let $V=(a,b)\times M$,
$g=-dt^{2}+h$ be a product spacetime. Then:
\begin{itemize}
\item[(i)] The lines of the boundary based on points in
$\partial_{B}(M)\setminus\partial_{C}(M)$ are null (i.e. any two
points on the line are horismotically related).
\item[(ii)] The lines of the boundary based on points in
$\partial_{C}(M)$ are timelike (i.e. any two points on the line are
chronologically related).
\item[(iii)] The copies of $M$ in the boundary are spacelike (i.e. any two points on the copy are not causally related).
\end{itemize}
\end{proposition}
{\it Proof.} {\it (i)} Let $(P,\emptyset)$, $(P',\emptyset)$ be two
elements of the boundary lying on some line based on some point in
$\partial_{B}(M)\setminus\partial_{C}(M)$. Then, it is not a
restriction to assume
$$P=\{(t',x'): t'<b_{c}(x')\},\quad P'=\{(t',x'): t'<b_{c'}(x')\},\quad\hbox{being}\; b_{c'}=b_{c}+k,\; k>0.$$ Therefore, $P\varsubsetneq P'$,
which proves that $(P,\emptyset)$ and $(P',\emptyset)$ are causally
related. On the other hand, the equality $\emptyset\cap
P'=\emptyset$ implies that they are not chronologically related.
Whence $(P,\emptyset)$ and $(P',\emptyset)$ are horismotically
related.

{\it (ii)} Let $(P,F)$, $(P',F')$ be two elements of the boundary
lying on some line based on some point $x_{0}\in\partial_{C}(M)$. We
can assume that
\[
F=\{(t',x'): t'>-d^{-\Omega}_{x_{0}}(x')\},\;\; P'=\{(t',x'):
t'<d^{\Omega'}_{x_{0}}(x')\},\;\;\hbox{being}\;\,
a\leq\Omega<\Omega'\leq b.
\]
As consequence, if $\overline{\Omega}=\Omega+(\Omega'-\Omega)/2$ and
$x\in M$ is such that $d(x,x_{0})<(\Omega'-\Omega)/2$, then
$\overline{\Omega}>-d_{x_{0}}^{-\Omega}(x)$,
$\overline{\Omega}<d_{x_{0}}^{\Omega'}(x)$, and thus,
$(\overline{\Omega},x)\in P'\cap F\neq\emptyset$. Whence, $(P,F)$
and $(P',F')$ are chronologically related.

{\it (iii)} For example, let $(P,\emptyset)\neq (P',\emptyset)$ be
two elements of some copy of $M$ in the boundary. Then
$P=I^{-}[\gamma_{+}]$ for some inextensible timelike curve
$\gamma_{+}:[w,b)\rightarrow V$, $\gamma_{+}(t)=(t,c_{+}(t))$,
$c_{+}(t)\rightarrow x_{0}\in M$, and $P'=I^{-}[\gamma'_{+}]$ for
some inextensible timelike curve $\gamma'_{+}:[w,b)\rightarrow V$,
$\gamma'_{+}(t)=(t,c'_{+}(t))$, $c'_{+}(t)\rightarrow x'_{0}\in M$,
$x_{0}\neq x'_{0}$. In particular, $t>d_{x_{0}}^{b}(c'_{+}(t))$,
$t>d^{b}_{x'_{0}}(c_{+}(t))$ for all $t$ close enough to $b<\infty$.
Whence, $P\nsubseteq P'$, $P'\nsubseteq P$, which implies that
$(P,\emptyset)$, $(P',\emptyset)$ are not causally related. $\Box$

\vspace{2mm}

Finally, the topology for the causal completion is directly deduced
from the definition of limit operator $L$ in terms of $\hat{L}$ and
$\check{L}$ (formulae (\ref{ee})):
\begin{proposition} The chronological topology on $\overline{V}$ coincides with the quotient topology under $\sim_{S}$ of the topology generated
by the limit operators $\hat{L}$ and $\check{L}$ on $\hat{V}\cup
\check{V}$.
\end{proposition}

\vspace{3mm}

\noindent \textbf{\em $\S$6. Main result:}

\vspace{2mm}

All these propositions, joined to the following result from
\cite[Prop. 6.7, Sect. 6.1.3]{FH}, yield our main statement, Theorem
\ref{central}.
\begin{proposition}\label{aa} Let $M=(\alpha,\omega)\times_{a}K$ be a Riemannian manifold with $(\alpha,\omega)\subseteq\R$, $a:(\alpha,\omega)\rightarrow\R$
a positive function, $K$ a compact manifold, and metric given by
$h=d\rho^{2}+a(\rho)^{2}j_{K}$ (being $j_{k}$ the metric on $K$).
For some $\rho_{-}<\rho_{+}$ in $(\alpha,\omega)$, assume that
$a(\rho)$ is decreasing in $\rho\in (\alpha,\rho_{-}]$ and
increasing in $\rho\in [\rho_{+},\omega)$, or increasing in the
whole $(\alpha,\omega)$ with $a(\alpha)=0$, $\alpha=0$. Then
$\partial_{B}(M)$ is formed by two spaces $B_{\alpha}$ and
$B_{\omega}$, attached at $\{\alpha\}\times K$ and
$\{\omega\}\times K$, resp., with each $B_{\imath}$
($\imath=\alpha,\omega$) being $K$ or an unique point $\ast$:
concretely, $B_{\imath}\cong K$ if
$|\int_{\rho_{0}}^{\imath}1/a(\rho)^{2}d\rho|<\infty$, and
$B_{\imath}=\ast$ if
$|\int_{\rho_{0}}^{\imath}1/a(\rho)^{2}d\rho|=\infty$. Moreover,
$B_{\imath}$ belongs to $\partial_{C}(M)$ if and only if the
extreme $\imath$ is finite.
\end{proposition}

\vspace{1mm}

\begin{theorem}\label{central} Let $V=(a,b)\times M$, $g=-dt^{2}+h$ be
a product spacetime whose spatial part $(M,h)$ falls under the
hypotheses of Proposition \ref{aa}. Then, the causal boundary of
$(V,g)$ admits the following structure (with the chronological
topology):
\begin{itemize}
\item[(i)] If $-\infty=a<b=\infty$ then it is formed by two
infinity null cones, one for the future and another for the past,
with base $\partial_{B}(M)\setminus\partial_{C}(M)$ and apexes
$i^{+}$ and $i^{-}$, resp., and timelike lines of future and past
extremes $i^{+}$ and $i^{-}$, resp., on each point of
$\partial_{C}(M)$ (Figure 1). \item[(ii)] If $-\infty<a<b<\infty$
then it is formed by two copies, one for the future and another for
the past, of the Cauchy completion $\overline{M}_{C}$ of $(M,h)$,
and timelike lines based on each point of $\partial_{C}(M)$ which
connect both copies (Figure 2).
\item[(iii)] If $-\infty<a<b=\infty$ then it is formed by an
infinity null cone for the future with base
$\partial_{B}(M)\setminus\partial_{C}(M)$ and apex $i^{+}$, a copy
of $\overline{M}_{C}$ for the past, and timelike lines based on each
point of $\partial_{C}(M)$ which connect $\overline{M}_{C}$ with
$i^{+}$ (Figure 3). \item[(iv)] If $-\infty=a<b<\infty$ then it is
formed by an infinity null cone for the past with base
$\partial_{B}(M)\setminus\partial_{C}(M)$ and apex $i^{-}$, a copy
of $\overline{M}_{C}$ for the future, and timelike lines based on
each point of $\partial_{C}(M)$ which connect $i^{-}$ with
$\overline{M}_{C}$ (Figure 4).
\end{itemize}
Moreover, $\partial_{B}(M)$ is formed by two spaces $B_{\alpha}$
and $B_{\omega}$, attached at $\{\alpha\}\times K$ and
$\{\omega\}\times K$, resp., with each $B_{\imath}$
($\imath=\alpha,\omega$) being $K$ or an unique point $\ast$:
concretely, $B_{\imath}\cong K$ if
$|\int_{\rho_{0}}^{\imath}1/a(\rho)^{2}d\rho|<\infty$, and
$B_{\imath}=\ast$ if
$|\int_{\rho_{0}}^{\imath}1/a(\rho)^{2}d\rho|=\infty$. Finally,
$B_{\imath}$ belongs to the Cauchy boundary $\partial_{C}(M)$ if
and only if the extreme $\imath$ is finite.
\end{theorem}

In particular, Theorem \ref{central} shows that the causal boundary
of a product spacetime is exclusively determined by the Busemann
boundary $\partial_{B}(M)$ (and the Cauchy boundary
$\partial_{C}(M)$) of the spatial part $(M,h)$ and the extremes of
the temporal interval $I$.

\begin{remark}\label{obs} {\em If the spatial part $(M,h)$ does not fall under the hypotheses of Proposition \ref{aa}, the structure statements of the first part of Theorem \ref{central} are still true.}
\end{remark}

\begin{figure}[ht]
\begin{center}
\includegraphics[width=5.2cm]{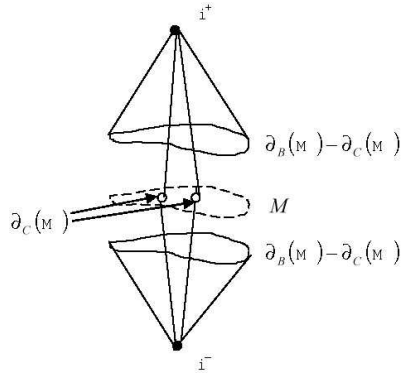}
\caption{Causal boundary for product spacetime with
$-\infty=a<b=\infty$.}
\end{center}
\end{figure}

\newpage

\begin{figure}[ht]
\begin{center}
\includegraphics[width=3.7cm]{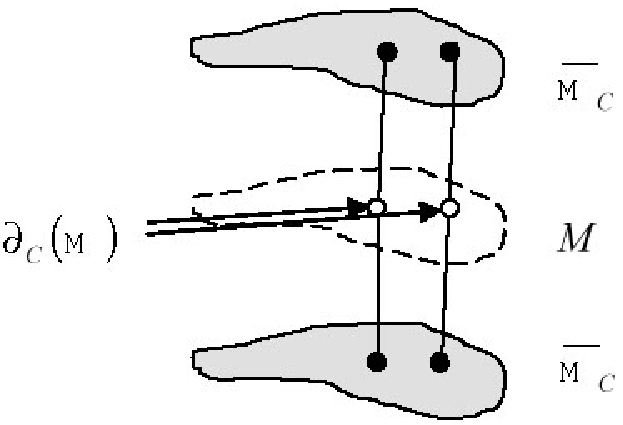}
\caption{Causal boundary for product spacetime with
$-\infty<a<b<\infty$.}
\end{center}
\end{figure}

\begin{figure}[ht]
\begin{center}
\includegraphics[width=5.2cm]{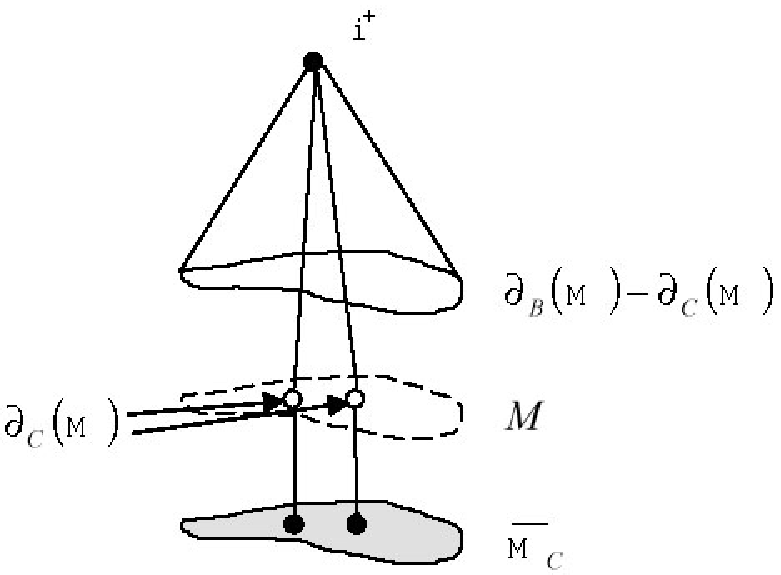}
\caption{Causal boundary for product spacetime with
$-\infty<a<b=\infty$.}
\end{center}
\end{figure}

\begin{figure}[ht]
\begin{center}
\includegraphics[width=3.7cm]{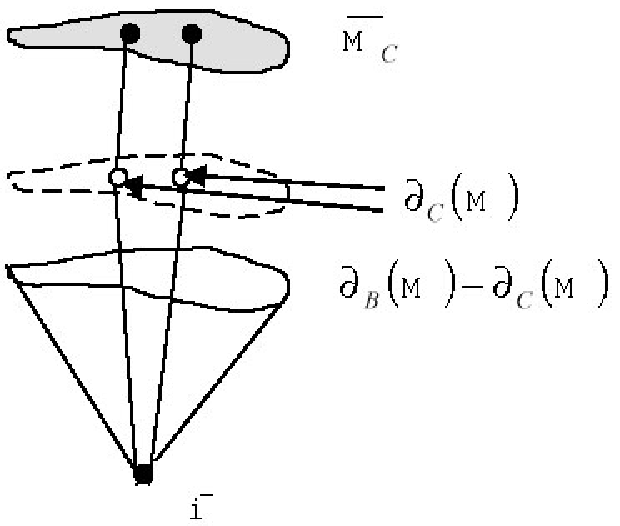}
\caption{Causal boundary for product spacetime with
$-\infty=a<b<\infty$.}
\end{center}
\end{figure}

\section{Applications}\label{examples}\label{4}

Even though product spacetimes present a relatively simple
structure, the conformal class is very general, including families
of spacetimes of great interest in Relativity. Concretely, we have
{\em (standard) static spacetimes} and {\em Generalized
Robertson-Walker spacetimes}.

\subsection{Static spacetimes}

(Standard) static spacetimes can be written as
\begin{equation}\label{e22}
V=\R\times M,\qquad g=-\beta(x)dt^{2}+g_{0},
\end{equation}
where $(M,g_{0})$ is a Riemannian manifold and $\beta$ a positive
function defined on $M$. A systematic study of the partial
boundaries for these spacetimes was initiated by Harris in
\cite{H3}, and then continued in collaboration with the second
author in \cite{FH}. However, these papers do not deal with the
question of how to attach the partial boundaries together in order
to form the (total) causal boundary. In this section we are going to
apply Theorem \ref{central} to cover this deficiency.

First, apply a conformal transformation to (\ref{e22}), with
conformal factor $f(x)=1/\beta(x)$. We obtain the new metric
\[
\overline{g}:=f(x)\cdot
g=-dt^{2}+\frac{g_{0}}{\beta(x)}=-dt^{2}+h,\qquad\hbox{where}\;\;
h:=g_{0}/\beta.
\]
So, the conformal invariance of the causal boundary reduces the
problem to study the product spacetime:
\[
V=\R\times M,\qquad g=-dt^{2}+h,
\]
where we have renamed $\overline{g}$ by $g$. This spacetime falls
under the hypothesis {\it (i)} of Theorem \ref{central}. Therefore:
\begin{theorem}\label{princ} Let $(V,g)$ be a static spacetime as in (\ref{e22}). Assume that the spatial part
$M$ endowed with metric $h=g_{0}/\beta$ falls under the hypotheses
of Proposition \ref{aa}. Then, the causal boundary (with the
chronological topology) is formed by two null cones, one for the
future and another for the past, with base
$\partial_{B}(M)\setminus\partial_{C}(M)$ and apexes $i^{+}$ and
$i^{-}$, resp., and timelike lines on each element of
$\partial_{C}(M)$, with future and past extremes $i^{+}$ and
$i^{-}$, resp. (Figure 1). Moreover, $\partial_{B}(M)$ is formed by
two spaces $B_{\alpha}$ and $B_{\omega}$, attached at
$\{\alpha\}\times K$ and $\{\omega\}\times K$, resp., with each
$B_{\imath}$ ($\imath=\alpha,\omega$) being $K$ or an unique point
$\ast$: $B_{\imath}\cong K$ if
$|\int_{\rho_{0}}^{\imath}1/a(\rho)^{2}d\rho|<\infty$, and
$B_{\imath}=\ast$ if
$|\int_{\rho_{0}}^{\imath}1/a(\rho)^{2}d\rho|=\infty$. Finally,
$B_{\imath}$ belongs to the Cauchy boundary $\partial_{C}(M)$ if and
only if the extreme $\imath$ is finite.
\end{theorem}

\vspace{1mm}

\noindent Next, we are going to apply this result to some
classical static spacetimes:

\vspace{3mm}

\noindent $\bullet$ {\bf Reissner-Nordstrom spacetime:}

\vspace{2mm}

In local coordinates, this spacetime can be written as:
\[
V=\R\times (0,\infty)\times \e^{2},\qquad
g=-f(r)dt^{2}+f(r)^{-1}dr^{2}+r^{2}(d\theta^{2}+\sin^{2}\theta
d\phi^{2}),
\]
where
\[
f(r):=1-\frac{2m}{r}+\frac{q^{2}}{r^{2}}.
\]
It models the gravitational field outside an electrically charged
massive object which is spherically symmetric. The constants $m>0$
and $q$ can be identified with the gravitational mass and the
electric charge of the object. The static regions of this spacetime
are determined by condition $f(r)>0$. So, we distinguish two cases:

\vspace{2mm}

-{\em Case weakly or critically charged,} $|q|\leq 2m$: Here, $f(r)$
is positive for $r\in (0,r^{-})\cup (r^{+},\infty)$, where
$r^{\pm}=m(1\pm\sqrt{1-(q/m)^{2}})$. First, consider the exterior
region $V=\R\times (r^{+},\infty)\times \e^{2}$. We have
\[
M=(r^{+},\infty)\times \e^{2},\quad
g_{0}=f(r)^{-1}dr^{2}+r^{2}(d\theta^{2}+\sin^{2}\theta
d\phi^{2})\quad\hbox{y}\quad \beta(x)=f(r).
\]
Observe that $M$ endowed with metric $h=g_{0}/\beta$ falls under
the hypotheses of Proposition \ref{aa}, being
\[
(K,j_{K})=(\e^{2},d\theta^{2}+\sin^{2}\theta d\phi^{2}),\quad
a(\rho)=\sqrt{r^{2}/f(r)},
\]
and
\[
\alpha=\int^{r^{+}}_{r_{0}^{+}}1/f(r)dr=-\infty,\qquad
\omega=\int_{r_{0}^{+}}^{\infty}1/f(r)dr=\infty,\qquad r^{+}_{0}\in
(r^{+},\infty)
\]
Moreover,
\[
\left|\int_{\rho_{0}=0}^{\alpha=-\infty}\frac{1}{a(\rho)^{2}}d\rho\right|=\int_{r^{+}}^{r_{0}^{+}}\frac{dr}{r^{2}}<\infty,\qquad
\left|\int_{\rho_{0}=0}^{\omega=\infty}\frac{1}{a(\rho)^{2}}d\rho\right|=\int_{r_{0}^{+}}^{\infty}\frac{dr}{r^{2}}<\infty.
\]
Therefore: {\em the causal boundary of the exterior region of weakly
or critically charged Reissner-Nordstrom (with the chronological
topology) is formed by two null cones at $r=r_{+}$ with base
$\e^{2}$ and apexes $i^{+}$, $i^{-}$, and two null cones at infinity
with base $\e^{2}$ and the same apexes $i^{+}$, $i^{-}$ (Figure 5).
In particular, this is also the causal boundary for the exterior
region of Schwarzschild spacetime ($q=0$).}

This is in agreement with the conformal approach. The double cone at
$r=r_{+}$ is due to the fact that only the region $r>r_{+}$ is
considered. On the other hand, the double cone at infinity is due to
the similarity between Reissner-Nordstrom and Minkowski far away
from the source.

\vspace{1mm}

Consider now the interior region $V=\R\times (0,r^{-})\times
\e^{2}$. We have
\[
M=(0,r^{-})\times \e^{2},\quad
g_{0}=f(r)^{-1}dr^{2}+r^{2}(d\theta^{2}+\sin^{2}\theta
d\phi^{2})\quad\hbox{and}\quad \beta(x)=f(r).
\]
The spatial part $M$ endowed with metric $h=g_{0}/\beta$ falls
again under the hypotheses of Proposition \ref{aa}, with
$(K,j_{K})$ and $a(\rho)$ as before, but now
\[
\alpha=\int_{r^{-}_{0}}^{0}1/f(r)dr>-\infty,\qquad
\omega=\int_{r_{0}^{-}}^{r^{-}}1/f(r)dr=\infty,\qquad r^{-}_{0}\in
(0,r^{-}).
\]
Moreover,
\[
\left|\int_{\rho_{0}=0}^{\alpha}\frac{1}{a(\rho)^{2}}d\rho\right|=\int_{0}^{r_{0}^{-}}\frac{dr}{r^{2}}=\infty,\qquad
\left|\int_{\rho_{0}=0}^{\omega=\infty}\frac{1}{a(\rho)^{2}}d\rho\right|=\int_{r_{0}^{-}}^{r^{-}}\frac{dr}{r^{2}}<\infty.
\]
Therefore: {\em the causal boundary of the interior region of weakly
or critically charged Reissner-Nordstrom (with the chronological
topology) is formed by two null cones at $r=r^{-}$, with base
$\e^{2}$ and apexes $j^{+}$ and $j^{-}$, joined by both extremes to
an unique timelike line at $r=0$, the central singularity (Figure
5).}

This result justifies rigourously the identifications between the
future and past timelike lines at $r=0$ suggested in \cite[Sect.
6.1.3]{FH} `without proof'. Notice also that the Reissner-Nordstrom
singularity becomes $1$-dimensional, in contraposition to the
well-known $\R\times \e^{2}$ structure of the Schwarzschild
singularity.

\begin{figure}[ht]
\begin{center}
\includegraphics[width=11cm]{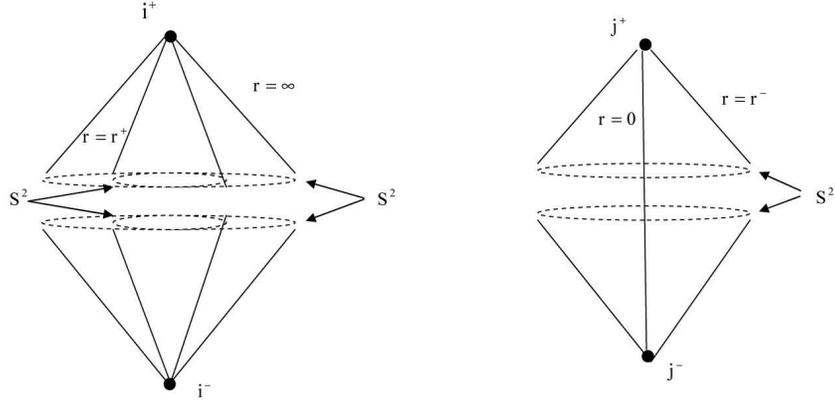}
\caption{Causal boundary of exterior region (left picture) and
interior region (right picture) of weakly or critically charged
Reissner-Nordstrom.}
\end{center}
\end{figure}

\vspace{2mm}

-{\em Case strongly charged,} $|q|>2m$: Now, $f(r)>0$ for all $r$,
and so, the static region coincides with the whole spacetime
$V=\R\times (0,\infty)\times\e^{2}$. Therefore, $M=(0,\infty)\times
\e^{2}$, and $g_{0}$, $\beta$ maintain the same expression as
before. The spatial part $M$ endowed with metric $h=g_{0}/\beta$
falls under the hypotheses of Proposition \ref{aa}, being
$(K,j_{K})$ and $a(\rho)$ as in previous cases, and
$$\alpha=\int^{0}_{r_{0}}1/f(r)dr>-\infty,\qquad\omega=\int_{r_{0}}^{\infty}1/f(r)dr=\infty.$$
Moreover,
\[
\left|\int_{\rho_{0}=0}^{\alpha}\frac{1}{a(\rho)^{2}}d\rho\right|=\int_{0}^{r_{0}}\frac{dr}{r^{2}}=\infty,\qquad
\left|\int_{\rho_{0}=0}^{\omega=\infty}\frac{1}{a(\rho)^{2}}d\rho\right|=\int_{r_{0}}^{\infty}\frac{dr}{r^{2}}<\infty.
\]
Therefore: {\em the causal boundary of strongly charged
Reissner-Nordstrom (with the chronological topology) is formed by
two null cones at $r=\infty$ with base $\e^{2}$ and apexes $i^{+}$,
$i^{-}$, and a timelike line at $r=0$ with the same extremes
$i^{+}$, $i^{-}$. (In this case, the diagram corresponds to the
right picture in Figure 5 with $r^{-}=\infty$ and $j^{+}, j^{-}$
replaced by $i^{+}, i^{-}$.)}

\vspace{3mm}

\noindent $\bullet$ {\bf Anti-de Sitter spacetime:}

\vspace{2mm}

This spacetime of constant sectional curvature $-1$ and topology
$\e^{1}\times \R^{3}$ does contain closed timelike curves. In
particular, it is not strongly causal, and thus, the causal boundary
approach cannot apply. However, this spacetime is not $1$-connected,
being its universal cover strongly causal and static. Consequently,
in this section by {\em Anti-de Sitter spacetime} we will understand
its universal cover.

In local coordinates, Anti-de Sitter spacetime can be written as
\[
V=\R\times (0,\infty)\times\e^{2},\qquad
g=-\cosh^{2}(r)dt^{2}+dr^{2}+\sinh^{2}(r)(d\theta^{2}+\sin^{2}\theta
d\phi^{2}).
\]
According to the notation previously introduced, we have
\[
M=(0,\infty)\times \e^{2},\quad
g_{0}=dr^{2}+\sinh^{2}(r)(d\theta^{2}+\sin^{2}\theta
d\phi^{2})\quad\hbox{y}\quad \beta(x)=\cosh^{2}(r).
\]
The spatial part $M$ endowed with metric $h=g_{0}/\beta$ falls
under the hypotheses of Proposition \ref{aa}, being
\[
(K,j_{K})=(\e^{2},d\theta^{2}+\sin^{2}\theta d\phi^{2}),\quad
a(\rho)=\tanh(r),
\]
and
\[
\alpha=\int_{r_{0}=1}^{0}\frac{dr}{\beta(x)}=\int_{1}^{0}\frac{dr}{\cosh(r)}>-\infty,\qquad
\omega=\int_{r_{0}=1}^{\infty}\frac{dr}{\beta(x)}=\int_{1}^{\infty}\frac{dr}{\cosh(r)}dr<\infty.
\]
Moreover,
\[
\left|\int^{\alpha}_{\rho_{0}=0}\frac{1}{a(\rho)^{2}}d\rho\right|=\int_{0}^{1}\frac{\cosh(r)}{\sinh^{2}(r)}dr=\infty,\quad
\left|\int_{\rho_{0}=0}^{\omega}\frac{1}{a(\rho)^{2}}d\rho\right|=\int_{1}^{\infty}\frac{\cosh(r)}{\sinh^{2}(r)}dr<\infty.
\]
Therefore, if we ignore the boundary region associated to $r=0$,
which is not representative because it corresponds to a coordinate
singularity, we obtain the following structure: {\em the causal
boundary of Anti-de Sitter spacetime (with the chronological
topology) is formed just by a timelike surface at $r=\infty$ of
section $\e^{2}$, which connects the points $i^{+}$ and $i^{-}$
(Figure 6).}

\vspace{1mm}

In particular, this structure coincides with the conformal boundary.
Notice however that points $i^{+}$, $i^{-}$ are isolated in the
conformal approach \cite[p. 132]{HE}.

\begin{figure}[ht]
\begin{center}
\includegraphics[width=4.7cm]{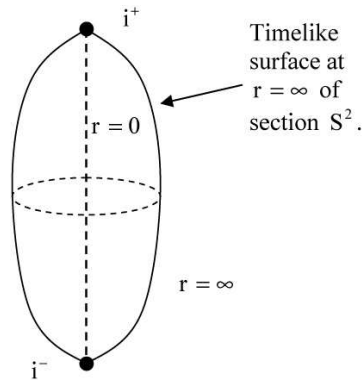}
\caption{Causal boundary of Anti-de Sitter spacetime.}
\end{center}
\end{figure}

\vspace{3mm}

\noindent $\bullet$ {\bf Taub spacetime:}

\vspace{2mm}

This spacetime, firstly introduced by A. H. Taub in \cite{T}, can
be written as:
\[
V=\R^{3}\times (0,\infty),\qquad
g=\rho^{-1/2}(-dt^{2}+d\rho^{2})+\rho(dx^{2}+dy^{2}),\;\;
(\rho>0).
\]
It is the unique solution of empty Einstein equations without
cosmological term, which is plane symmetric and static. The causal
boundary of Taub spacetime was studied by Kuang, Li and Liang in
\cite{KLL}. Surprisingly, they found that the singular region of the
boundary reduces to one point when the GKP construction is applied,
in contraposition to the $1$-dimensional structure physically
expected. Since then, this result has been claimed as an important
evidence that GKP approach is not totally satisfactory. Next, we are
going to apply Theorem \ref{princ} in order to show that this defect
is not present in the approach followed in this paper.

According to the notation previously introduced, now we have:
\[
M=(0,\infty)\times \R^{2},\quad g_{0}=\rho^{-1/2}d\rho^{2}+\rho
dx^{2}+\rho dy^{2}\quad\hbox{y}\quad \beta(x)=\rho^{-1/2}.
\]
Therefore,
\[
(K,j_{K})=(\R^{2},dx^{2}+dy^{2}),\quad
(\alpha,\omega)=(0,\infty)\quad\hbox{y}\quad a(\rho)=\rho^{3/2}.
\]
In this case, the spatial part $M$ endowed with metric
$h=g_{0}/\beta$ does not fall under the hypotheses of Proposition
\ref{aa}, since $K=\R^{2}$ is not compact. However, from the
analysis and classification of the pasts and futures of inextensible
causal curves developed in \cite[pp. 1534-5]{KLL}, it implicitly
follows
\[
\partial_{B}(M)=B_{\alpha=0}\cup B_{\omega=\infty},
\quad\hbox{being}\;\; B_{0}=*,\;\; B_{\infty}=\R^{2}.
\]
Therefore, from the first part of Theorem \ref{princ} (recall Remark
\ref{obs}), we obtain: {\em the causal boundary of Taub spacetime is
formed by a timelike line at $\rho=0$ of future and past extremes
$i^{+}$, $i^{-}$, resp., and two null cones at $\rho=\infty$, one
for the future and another for the past, with base $\R^{2}$ and
apexes $i^{+}$, $i^{-}$, resp. (Figure 7).}

\vspace{1mm}

In conclusion, this result does reproduce the $1$-dimensional
character of the Taub singularity, represented by the region of the
boundary at $\rho=0$.

\begin{figure}[ht]
\begin{center}
\includegraphics[width=4.7cm]{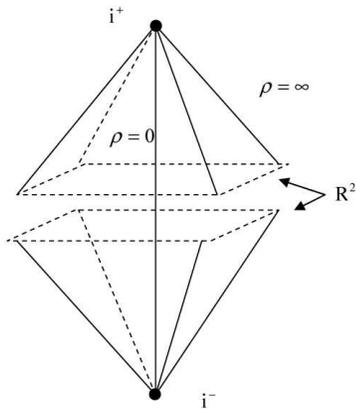}
\caption{Causal boundary of Taub spacetime.}
\end{center}
\end{figure}

\subsection{Generalized Robertson-Walker spacetimes}

By Generalized Robertson-Walker (GRW) spacetimes we understand the
family of spacetimes given by:
\begin{equation}\label{met-alab}
V=(a',b')\times M,\qquad g=-dt^{2}+\alpha(t)h,
\end{equation}
where $(a',b')\subseteq\R$ is an open interval of $\R$ called {\em
base}, $(M,h)$ is an arbitrary Riemannian manifold called {\em
fiber} and $\alpha(t)$ is a positive function defined on $(a',b')$
called {\em warping function} or {\em scale factor}. This family is
quite important in Relativity, since it provides a first approach to
the global structure of the universe: not for nothing it arises as a
natural generalization of the standard cosmological models (studied
in a moment). See \cite{Sa,GB} for local and global geometrical
characterizations of GRW spacetimes.

The causal boundary of these spacetimes can be obtained by applying
the result \cite[Proposition 5.2]{H2}, which describes the partial
boundaries for the more general family of {\em multiwarped
spacetimes} (i.e., multiple fibers and multiple warping functions
considered). However, this result does not work when the resulting
boundary has non-spacelike regions. In this section we are going to
use Theorem \ref{central} to extend this result to cover any causal
character for the boundary, at least in the smaller class of GRW
spacetimes.

To this aim, first apply a conformal transformation to
(\ref{met-alab}), with conformal factor $f(t)=1/\alpha(t)$. We
obtain the new metric
\[
\overline{g}:=f(t)\cdot g=-\frac{dt^{2}}{\alpha(t)}+h=-ds^{2}+h,
\]
where the variable $s$ is defined by the relation
$ds=dt/\sqrt{\alpha(t)}$. Taking into account the conformal
invariance of the causal boundary, we only need to study the product
spacetime
\[
V=(a,b)\times M,\qquad \overline{g}=-ds^{2}+h,
\]
where $I=(a,b)$ is the domain for the new variable $s$, and thus, it
may be different from the initial interval $(a',b')$. From the
relation between $s$ and $t$ it directly follows:
\[
a=-\int^{c_{0}}_{a'}\frac{1}{\sqrt{\alpha(t)}}dt,\qquad
b=\int^{b'}_{c_{0}}\frac{1}{\sqrt{\alpha(t)}}dt,\qquad
a'<c_{0}<b'.
\]
Therefore, we conclude that the causal boundary of GRW spacetime
(\ref{met-alab}) coincides with that of the spacetime
\begin{equation}\label{t1}
V=I\times M,\qquad g=-dt^{2}+h,
\end{equation}
where we have renamed $\overline{g}$ and $s$ by $g$ and $t$, resp.
Theorem \ref{central} applied to (\ref{t1}) then provides the
following result:
\begin{theorem}\label{central-RWG} Let $V=(a',b')\times
M$, $g=-dt^{2}+\alpha(t)h$ be a GRW spacetime with spatial part
$(M,h)$ under the hypotheses of Proposition \ref{aa}. Then, the
causal boundary (with the chronological topology) has the following
structure:
\begin{itemize}
\item[(i)] If
$\int_{a'}^{c_{0}}1/\sqrt{\alpha(t)}dt=\int^{b'}_{c_{0}}1/\sqrt{\alpha(t)}dt=\infty$
then it is formed by two infinity null cones, one for the future and
another for the past, with base
$\partial_{B}(M)\setminus\partial_{C}(M)$ and apexes $i^{+}$ and
$i^{-}$, resp., and timelike lines of future and past extremes
$i^{+}$ and $i^{-}$, resp., on each point of $\partial_{C}(M)$
(Figure 1). \item[(ii)] If $\int_{a'}^{c_{0}}1/\sqrt{\alpha(t)}dt,\,
\int^{b'}_{c_{0}}1/\sqrt{\alpha(t)}dt<\infty$ then it is formed by
two copies, one for the future and another for the past, of the
Cauchy completion $\overline{M}_{C}$ of $(M,h)$, and timelike lines
based on each point of $\partial_{C}(M)$ which connect both copies
(Figure 2). \item[(iii)] If
$\int_{a'}^{c_{0}}1/\sqrt{\alpha(t)}dt<\infty$,
$\int^{b'}_{c_{0}}1/\sqrt{\alpha(t)}dt=\infty$ then it is formed by
an infinity null cone for the future with base
$\partial_{B}(M)\setminus\partial_{C}(M)$ and apex $i^{+}$, a copy
of $\overline{M}_{C}$ for the past, and timelike lines based on each
point of $\partial_{C}(M)$ which connect $\overline{M}_{C}$ with
$i^{+}$ (Figure 3). \item[(iv)] If
$\int_{a'}^{c_{0}}1/\sqrt{\alpha(t)}dt=\infty$,
$\int^{b'}_{c_{0}}1/\sqrt{\alpha(t)}dt<\infty$ then it is formed by
an infinity null cone for the past with base
$\partial_{B}(M)\setminus\partial_{C}(M)$ and apex $i^{-}$, a copy
of $\overline{M}_{C}$ for the future, and timelike lines based on
each point of $\partial_{C}(M)$ which connect $i^{-}$ with
$\overline{M}_{C}$ (Figure 4).
\end{itemize}
Moreover, $\partial_{B}(M)$ is formed by two spaces $B_{\alpha}$
and $B_{\omega}$, attached at $\{\alpha\}\times K$ and
$\{\omega\}\times K$, resp., with each $B_{\imath}$
($\imath=\alpha,\omega$) being $K$ or an unique point $\ast$:
concretely, $B_{\imath}\cong K$ if
$|\int_{\rho_{0}}^{\imath}1/a(\rho)^{2}d\rho|<\infty$, and
$B_{\imath}=\ast$ if
$|\int_{\rho_{0}}^{\imath}1/a(\rho)^{2}d\rho|=\infty$. Finally,
$B_{\imath}$ belongs to the Cauchy boundary $\partial_{C}(M)$ if
and only if the extreme $\imath$ is finite.
\end{theorem}

In particular, the structure of the causal boundary for GRW
spacetimes depends on both, the spatial part $(M,h)$ and the scale
factor $\alpha(t)$. If the spatial part $(M,h)$ is complete
($\partial_{C}(M)=\emptyset$), the boundary presents at each extreme
of the temporal interval $(a',b')$, either a spacelike cover
structure $M$ or a null cone with base $\partial_{B}(M)$, depending
on the growth of the scale factor at each extreme. However, if the
spatial part $(M,h)$ is incomplete ($\partial_{C}(M)\neq\emptyset$),
the boundary will also contain timelike lines on each point of
$\partial_{C}(M)$, reachable by observers of the universe in finite
proper time. Hence, the boundary is not necessarily time symmetric
and may contain regions of any causal character (null, timelike or
spacelike).

\vspace{3mm}

\noindent $\bullet$ {\bf FLRW spacetimes:}

\vspace{2mm}

By completeness, we are going to particularize previous result to
Friedman-Lemaitre-Robertson-Walker (FLRW) spacetimes, i.e. the
spatial part is now a geometric model. In this case, the spacetime
manifold $V_{k}$ is either $(a',b')\times\R^{3}$ if $k=0,-1$ or
$(a',b')\times\e^{3}$ if $k=1$. In local coordinates, the line
element reads
\[
g=-dt^{2}+\alpha(t)[d\rho^{2}+a(\rho,k)^{2}(d\theta^{2}+\sin^{2}\theta
d\phi^{2})],
\]
where $\alpha(t)$ is the scale factor and
\[
a(\rho,k)=\left\{\begin{array}{ll} \sin\rho & \hbox{if}\;\; k=1 \\
\rho & \hbox{if}\;\; k=0 \\ \sinh\rho & \hbox{if}\;\; k=-1.
\end{array}\right.
\]
Therefore, Theorem \ref{central-RWG} gives:
\begin{theorem}\label{central-RW}
The causal boundary of a FLRW spacetime, $V=(a',b')\times M$,
$g=-dt^{2}+\alpha(t)h$ with $(M,h)\equiv\R^{3}, \e^{3}$ or $\H^{3}$,
has the following structure:
\begin{itemize}
\item[(i)] If
$\int_{a'}^{c_{0}}1/\sqrt{\alpha(t)}dt=\int^{b'}_{c_{0}}1/\sqrt{\alpha(t)}dt=\infty$
then it is formed by two infinity null cones, one for the future and
another for the past, with base $\e^{2}$ and apex $i^{+}$ and
$i^{-}$, resp., if $M=\R^{3},\H^{3}$, or it is formed just by
$i^{+}$, $i^{-}$ if $M=\e^{3}$. \item[(ii)] If
$\int_{a'}^{c_{0}}1/\sqrt{\alpha(t)}dt,\,
\int^{b'}_{c_{0}}1/\sqrt{\alpha(t)}dt<\infty$ then it is formed by
two spacelike copies, one for the future and another for the past,
of $\R^{3}$ if $M=\R^{3},\H^{3}$, or $\e^{3}$ if $M=\e^{3}$.
\item[(iii)] If
$\int_{a'}^{c_{0}}1/\sqrt{\alpha(t)}dt<\infty$,
$\int^{b'}_{c_{0}}1/\sqrt{\alpha(t)}dt=\infty$ then it is formed by
an infinity null cone for the future with base $\e^{2}$ and apex
$i^{+}$ and a copy of $\R^{3}$ for the past if $M=\R^{3},\H^{3}$, or
it is formed by $i^{+}$ for the future and a copy of $\e^{3}$ for
the past if $M=\e^{3}$.
\item[(iv)] If
$\int_{a'}^{c_{0}}1/\sqrt{\alpha(t)}dt=\infty$,
$\int^{b'}_{c_{0}}1/\sqrt{\alpha(t)}dt<\infty$ then it is formed by
an infinity null cone with base $\e^{2}$ and apex $i^{-}$ and a copy
of $\R^{3}$ for the future if $M=\R^{3},\H^{3}$, or it is formed by
$i^{-}$ for the past and a copy of $\e^{3}$ for the future if
$M=\e^{3}$.
\end{itemize}
\end{theorem}
{\it Proof.} If $k=0,-1$, the spatial part is
\[
M=(0,\infty)\times \e^{2},\quad
h=d\rho^{2}+a(\rho,k)^{2}(d\theta^{2}+\sin^{2}\theta
d\phi^{2}),\;\;
a(\rho,k)=\left\{\begin{array}{ll} \rho & \hbox{if}\;\; k=0 \\
\sinh\rho & \hbox{if}\;\; k=-1.
\end{array}\right.
\]
In particular, $(M,h)$ falls under the hypotheses of Proposition
\ref{aa}. Therefore, the conclusion directly follows from Theorem
\ref{central-RWG} and the integrals
\[
\left|\int^{0}_{\rho_{0}=1}\frac{1}{a(\rho,k)^{2}}d\rho\right|=\infty,\qquad
\left|\int_{\rho_{0}=1}^{\infty}\frac{1}{a(\rho,k)^{2}}d\rho\right|<\infty.
\]
(Obviously, we have ignored the boundary region at $\rho=0$.)

Assume now that $k=1$. In this case, $a(\rho,1)=\sin\rho$, and thus,
the spatial part does not fall under the hypotheses of Proposition
\ref{aa}. However, from  Remark \ref{obs} the first part of Theorem
\ref{central-RWG} still holds. Whence, the conclusion directly
follows from $\partial_{B}(\e^{3})=\partial_{C}(\e^{3})=\emptyset$
and $\overline{M}_{C}=M=\e^{3}$. $\Box$

\vspace{1mm}

An immediate application of this result gives the causal boundary
for {\em Einstein Static Universe}. This spacetime is a FLRW model
with base $\R$, fiber $(M,h)\equiv \e^{3}$ and scale factor
$\alpha(t)\equiv 1$. Therefore, taking into account that
$\int_{-\infty}^{0}dt=\int^{\infty}_{0}dt=\infty$: {\em the causal
boundary of ESU is formed by the points, $i^{+}$, $i^{-}$.} In this
case, the weak (indeed, null) asymptotic growth of the scale factor
implies degeneration of the boundary in two unique points. On the
opposite side we have {\em de Sitter spacetime}. This is a FLRW
model with base $\R$, fiber $(M,h)\equiv \e^{3}$ and scale factor
$\alpha(t)=\cosh(t)$. Therefore, taking into account that
$\int_{-\infty}^{0}1/\cosh(t)dt=\int_{0}^{\infty}1/\cosh(t)dt<\infty$:
{\em the causal boundary of de Sitter spacetime is formed by two
spacelike copies of $\e^{3}$, one for the past and another for the
future.} So, it is the strong asymptotic growth of the scale factor
$\alpha(t)=\cosh(t)$ which produces this big boundary.

\section{Conclusions}\label{5}

The main result in this paper is Theorem \ref{central}, which
describes the causal boundary for product spacetimes of a Lorentz
interval and a Riemannian manifold. The huge conformal class of
these spacetimes, which includes both, (standard) static spacetimes
and Generalized Robertson-Walker spacetimes, makes this result
specially useful to deduce the boundary of multiple classical
spacetimes. In particular, we have explicitly described the causal
boundary for Reissner-Nordstrom (including Schwarzschild), Anti-de
Sitter, Taub and standard cosmological models as de Sitter spacetime
and Einstein Static Universe.

In this paper we have used the formulation of the causal boundary
introduced in \cite{MR}, and modified later in \cite{F}. As
consequence, we have tested this new formulation for the classical
spacetimes previously cited. In particular, we have found that the
causal boundary essentially reproduces the structure of the
conformal boundary in these cases: right identifications between the
temporal lines of the boundary, expected dimensionality for the
singular regions, satisfactory topology for the completion...

This paper can be considered a very initial step in the ambitious
project of describing the causal boundary of spacetimes $V=I\times
M$ with metric $g=-dt^{2}+h_{t}$, where $h_{t}$ may depend on time.
As an indication of the importance and generality of this problem,
just recall that {\em any} globally hyperbolic spacetime admits this
decomposition, see \cite{BeS}. Another step within this program may
consist of completing the results about the boundary of multiwarped
spacetimes in \cite{H2}. These spacetimes are very interesting
because, apart from including certain regions of Schwarzschild and
Reissner-Nordstrom uncovered in this paper, they also include
Bianchi type IX spacetimes (as Kasner), spacetimes with internal
degrees of freedom attached at every point and multidimensional
inflationary models.

\end{document}